\documentclass[12pt]{article}
\usepackage{bm}
\usepackage{url}
\usepackage{amssymb}
\usepackage{amsmath}
\usepackage{hyperref}

\title{Random Field Quantization Method}
\author{G\'abor Helesfai \\ E-mail: \href{mailto:helesfaigabor@gmail.com}{helesfaigabor@gmail.com}}

\begin{document}

\maketitle

\begin{abstract}

Today it still remains a challenge whether quantum mechanics has an underlying statistical explanation or not. While there are and were a lot of models trying to explain 
quantum phenomena with statistical methods these all failed on certain levels. In this paper a method is proposed that is not only based on a classical statistical framework but it 
has an underlying physical model behind it and it can explain some of the basic characteristics of quantum mechanics. It will be shown that if look at the properties of a charged 
particle in a random electric field one can obtain the discrete energy values of the harmonic oscillator and the infinite potential well, and also gives a good qualitative 
description of the double-slit experiment and measurement theory. Also the side-effect of the model is the emergence of a constant with an action dimension.

\end{abstract}

\section{Introduction}

Since the discovery of quantum mechanics there were a lot of debates whether this theory describes basic reality or there is an underlying statistical explanation to all the 
phenomena that quantum mechanics predicts and explains. Albert Einstein was one of those who thought that there has to be such a theory. Unfortunately, though there were many 
who tried to construct such a model these eventually failed (a good summary of these can be found in \cite{Einsteins_dream} and \cite{stochastic_models}). While there were some 
qualitative results (e.g. explaining the Bell-experiment or the uncertainty principle - see for example \cite{unceartanty_relations}) these were unable to reproduce other results.\\
In this paper a new method is introduced which is aimed to deal with these problems. This model has two major differences from other approaches. \\
First it is based on an intuitive physical model which does not require extra input, unknown fields or the results of quantum mechanics (thus this model is different then 
for example the model presented in \cite{statistical_kinetics}).\\
Second - which in turn will be the consequence of the first property - the underlying process of quantum mechanics will be not based on stochastic processes but rather a 
random field.\\
In the first section the physical model will be introduced. In the second second section I will show how one can construct an equation for the probability density function 
(which - as it will be later shown - is actually a functional). After that the equation will be solved for some special cases which will be followed by the conclusions and 
outlook.

\section{The model}

Forget quantum mechanics for a minute and let us try to find a possible model which explains why the electron of a hydrogen atom does not fall into the proton. The results of 
classical electrodynamics is clear, the electron radiates energy - since its acceleration is not zero - thus losing momentum, which inevitably results in it falling into the 
proton.\\
There is a way out of this though, and this comes from a fact that what this model describes is a case when only the electric and magnetic fields of the proton and the electron 
are relevant, everything else can be neglected. But let us suppose that we consider the effects of the electric and magnetic fields of the surrounding protons and electrons. At 
the macroscopic scale these rarely need to be included since their effect of the motions of other bodies are practically negligible. But it is not trivial to state that this is 
also the case for small regions.\\
This will be the essence of the proposed model. To sum it up briefly it is based on the following premise:\\

\textit{Closed systems are always approximations and only on the macroscopic scale can one neglect the effects of the surrounding environment. In the quantum regime these effects 
have to be taken into consideration}.\\

In the following the nature of the effect of the environment will be explored and a proposal for a model will be suggested which will enable one to calculate the effects of the 
environment for some specific case.

\subsection{The random electromagnetic field}

From now on only electromagnetic interactions are considered.\\
Because there are a lot of charges even in the near vicinity of each particle it is practical to treat the effect of these as if the particle would be in a random 
electromagnetic field.\\
Let's consider the electric field first. Consider a point with coordinates $\underline{x}$ at time $t$. Let us index all charges which are to be treated as part of the 
environment by $n = 1,\ 2,\dots ,\ N$. The electric field for each charge at this $\underline{x}$ point in space at the given $t$ time will be 
$\underline{E}_1(\underline{x},t),\dots ,\underline{E}_N(\underline{x},t)$ thus the the total electric field will be 

$$\underline{E}(\underline{x},t)=\sum_{i=1}^N \underline{E}_i(\underline{x},t).$$

Since even at a small scale the electric field of many particles cannot be measured exactly we want to treat each $\underline{E}_i(\underline{x},t)$ as a random variable.
And if all $\underline{E}_i(\underline{x},t)$ are treated as random variable the sum of them will also be a random variable. Since this variable depends not only of time but also 
the coordinates it will be a random field. In the following one can show from a few assumptions that this random field has a normal distribution as a probability density 
function.\\
First it is fair to assume that the particles of the environment are mostly independent. Though each of them are in the light cone of the space-time point $(\underline{x},t)$ 
they are not necessarily in each others light cone. Second it is also fair to assume that all the distributions of $\underline{E}_i(\underline{x},t)$ have finite expectation 
values and variances which basically means that - since the magnitude of the electric field depends only on the distance - their movement is confined to a finite region. Applying 
these assumptions to the central limit theorem the probability density function of $\underline{E}(\underline{x},t)$ will be of the form

$$f(\underline{E}(\underline{x},t))\sim\exp\left(-\frac{1}{2}(A)^{-1}_{ij}(\underline{x},t)(E_i(\underline{x},t) - E^0_i(\underline{x},t))(E_j(\underline{x},t) - E^0_j(\underline{x},t))\right)$$

where $A_{ij}(\underline{x},t)$ are the elements of the covariance matrix and $E^0_i(\underline{x},t)$ are elements of the expectation value (here we have used the 
Einstein-convention and $i,j = 1,\ 2,\ 3$). What can be said about these quantities?\\
First we can presume that the expectation value is zero. Indeed if it would not be zero the experiments would show a constant electric field in all experiments.\\
From the independence of the $\underline{E}_i(\underline{x},t)$ fields it follows that there is not a distinguished point $\underline{x}$ or time $t$ or a direction where this 
distribution would behave in any different matter. This means that the covariance matrix is of the form $$A_{ij} = \frac{1}{K'}\delta_{ij}$$ where $K'$ does not depend on 
$\underline{x}$ or t. Thus the probability density for the electric field will take the form 

$$f(\underline{E}(\underline{x},t))\sim\exp\left(-\frac{K'}{2}\underline{E}^2(\underline{x},t)\right)$$

for each $\underline{x}$ and $t.$ One may notice that the exponent is proportional to the electric field energy density so let us write $K'\to K'\epsilon_0$ thus obtaining 

$$f(\underline{E}(\underline{x},t))\sim\exp\left(-\frac{K'\epsilon_0}{2}\underline{E}^2(\underline{x},t)\right)$$

($\epsilon_0$ is the vacuum permittivity). Note that during the calculation neither the value of the coordinate nor the time was specified thus for each $\underline{x}$ and $t$ the 
above thought process holds. This means that for each of these coordinates and time the corresponding probabilities will be independent, thus if the electric field is given in 
more than one point in the space-time the probability density will be of the form

\begin{eqnarray}
f(\underline{E})\sim\exp\left(-\frac{K'\epsilon_0}{2}\sum_{\underline{x},t}\underline{E}^2(\underline{x},t)\right)\label{electric_density}
\end{eqnarray}

where the meaning of the sum depends on the points chosen in the space-time. Specifically if the points are a subspace of space-time the probability density will be

$$f(\underline{E})\sim\exp\left(-\frac{K'\epsilon_0}{2}\int_U d^3xdt\underline{E}^2(\underline{x},t)\right)$$

where $U$ is the subspace involved.\\
What does this formula mean? Let us specify an electric field in $U$ - let this be $\underline{E}_0(\underline{x},t).$ The probability that the field in $U$ will be between 
$\underline{E}_0$ and $\underline{E}_0 + \underline{dE}$ is then given by the formula 

$$p(\underline{E}\in [\underline{E}_0, \underline{E}_0 + d\underline{E}])\sim\exp\left(-\frac{K'\epsilon_0}{2}\int_U d^3xdt\underline{E}_0^2(\underline{x},t)\right)\mathcal{D}^3E$$

where the $\mathcal{D}$ symbol means that this is actually a functional (since $\underline{E}$ depends on both the coordinates and time).\\

An interesting feature of this formula that since $\epsilon_0\underline{E}^2$ has dimensions of energy density, and the integral is with respect to all three coordinates and 
time the integral $\epsilon_0\int d^3xdt\underline{E}^2$ will have action dimension, thus the constant $1/K'$ has action dimension as well (note that if the region was not a 
subspace but e.g. a line this constant would have a different dimension).\\
With the same thought process one obtains a similar expression for the magnetic field:

$$f(\underline{B}(\underline{x},t))\sim\exp\left(-\frac{\tilde{K}'}{2\mu_0}\underline{B}^2(\underline{x},t)\right)$$

Now we will argue that $\tilde{K}' = K'.$ If we look at the combined probability density function

\begin{align}
f(\underline{E}(\underline{x},t))f(\underline{B}(\underline{x},t))\sim\exp\left(-\frac{\tilde{K}'}{2\mu_0}\underline{B}^2(\underline{x},t)-\frac{K'\epsilon_0}{2}\underline{E}^2(\underline{x},t)\right)
\end{align}

we will notice that if $\tilde{K}' = K'$ then the exponent will be the zero-zero component of the electromagnetic stress energy tensor. So if later we want to derive a covariant 
formula and we want the probability density function to be covariant then we must chose these constants to be equal. Conversely if these constants would differ the exponent couldn't 
be expressed as a special case of a covariant quantity.\\
To simplify the following calculations from now only the random electric field will be considered.

\section{Derivation of the equation}

\subsection{Formulating the problem}

Let us suppose that there is a charged particle moving in a constant electric field $\underline{E}_0$ and a random electric field $\underline{E}$ with a distribution 
density (\ref{electric_density}). The question is what can we say about the distribution density of the particles position and/or momentum?\\
The equations of motion are governed by the Lorentz-force:

\begin{eqnarray}
m\frac{d^2\underline{r}}{dt^2} = q(\underline{E}_0(\underline{r}(t)) + \underline{E}(\underline{r}(t),t))\label{Lorentz}
\end{eqnarray}

where $\underline{r}$ is the position of the charged particle and $q$ is its charge. Let us suppose that the random field is given - let us denote it $\underline{E}_1$. We then 
solve equation (\ref{Lorentz}) for $\underline{r}(t)$ - let us denote this $\underline{r}_1(t)$. If the problem is constrained in the $[t_0,t_1]$ time interval the probability 
for this trajectory will be 

$$p_1\sim\exp\left(-\frac{K'\epsilon_0}{2}\int_{t_0}^{t_1}\underline{E_1}^2(\underline{r}_1(t),t)dt\right)\mathcal{D}^3E.$$

Thus by cosing specific values of the random field we can map the probability density function of the particles trajectory. This means that we can define the probability density 
function of the trajectory by

\begin{eqnarray}
f(\underline{r}(t))\sim\exp\left(-\frac{K'\epsilon_0}{2}\int_{t_0}^{t_1}\left(\frac{m\frac{d^2\underline{r}}{dt^2} - q\underline{E}_0}{q}\right)^2(\underline{r}(t),t)dt\right)\label{path_density}
\end{eqnarray}

which is obtained by expressing $\underline{E}$ from (\ref{Lorentz}), substituting it into (\ref{electric_density}) and integrating the exponent from $t_0$ to $t_1.$\\
The meaning of this expression is the following. In a random field the path of a particle can be (almost) anything, but these paths have different probabilities. If we want to 
calculate the probability of a path $\underline{r}(t)$ we insert it into expression (\ref{path_density}).\\

Some important observations before we continue:

\begin{itemize}
\item Since in this case the total probability was not obtained via integration over both space and time the dimension of $1/K'$ is not action but action density (since the 
integral with respect to the space coordinates is missing).
\item The probability densities obtained here are all assigned to a path, thus this approach is very similar to a path integral approach.
\item If we want to be precise the probability density is actually a functional since it depends on a function.
\item The meaning of $f(\underline{r}(t))$ is the following: the probability that a path will be in the interval $[\underline{r}(t), \underline{r}(t) + d\underline{r}(t)]$ is 
$p\sim f(\underline{r}(t))d^3r$
\end{itemize}

The formula (\ref{path_density}) describes the probability density for a given path. What we really want is a differential equation for $f(\underline{r}(t))$ thus its properties 
can investigated in more detail. This differential equation will be derived in the next section.

\subsection{The differential equation}

What we want to do is derive a differential equation for the probability function (\ref{path_density}). To do this we will use the following trick. Let $$S = \int L dt$$ be the 
action of a Lagrangian $L = L(x,\dot{x})$ and let $$g = \exp(KS) = \exp(K\int Ldt)$$ Then using the expression $p = \partial S/\partial x$ and the Hamilton-Jacobi equation 
$\partial S/\partial t = -H$ we obtain

\begin{align}
\frac{\partial g}{\partial t} &= -KHg\label{Ham_jac_bas1}\\
\frac{\partial g}{\partial x} &= Kpg\label{Ham_jac_bas2}
\end{align}

where $H$ is the Hamiltonian derived from the Lagrangian $L$ and $p$ is the canonical momenta. If the Hamiltonian is a polynomial expression of the momenta then (\ref{Ham_jac_bas2}) 
can be used to express these in terms of the derivatives of $g$ and if we substitute these into (\ref{Ham_jac_bas1}) we obtain a differential equation.\\
In our case the probability density (\ref{path_density}) has a very similar expression, the only problem is that in (\ref{path_density}) the second derivative appears (as it is 
shown in e.g. \cite{higher_order_hamiltonian} it is non-trivial to transform these types of Lagrangians to ones that contain only first derivatives because there might be 
issues with the positivity of the energy or the canonical transformations). To cure this we will use the following trick. Let us introduce the probability density functional 

\begin{eqnarray}
\tilde{f}(\underline{x},\underline{p})\sim\exp\left(-\frac{K'\epsilon_0}{2q^2}\int_{t_0}^{t_1}\left[\left(\frac{d\underline{p}}{dt} - q\underline{E}_0(\underline{x}(t))\right)^2 + \lambda\left(m\frac{d\underline{x}}{dt} - \underline{p}\right)^2\right]\right)\label{path_density_mod}
\end{eqnarray}

Here $x$ and $p$ are two independent variables and $\lambda > 0$ is a currently undetermined parameter. This is also a probability density functional and the 
conditional probability density $$\tilde{f}(\underline{x},\underline{p}|\underline{p} - m\dot{\underline{x}} = 0)$$ will be proportional to the original probability density 
(\ref{path_density}). Thus if we work with (\ref{path_density_mod}) then we will obtain all information for the original probability density. Note that because the results need 
to be equal only "on-shell" - that is only when $\underline{p} - m\dot{\underline{x}} = 0$ - it is not necessary that the exponent of (\ref{path_density_mod}) and 
(\ref{path_density}) be related by any additional way. Specifically we will calculate the "Hamiltonian" for the exponent of (\ref{path_density_mod}) and it is not necessary 
that this is obtained via canonical transformation of the Hamiltonian calculated from the original exponent.\\

Now let the "Lagrangian" be

\begin{eqnarray}
\tilde{L} = \frac{1}{2}\left[\left(\frac{d\underline{p}}{dt} - q\underline{E}_0(\underline{x}(t))\right)^2 + \lambda\left(m\frac{d\underline{x}}{dt} - \underline{p}\right)^2\right]
\end{eqnarray}

and let us calculate the canonical moments of $\underline{x}$ and $\underline{p}:$

\begin{align}
\underline{P} &= \frac{\partial\tilde{L}}{\partial\underline{\dot{x}}} = \lambda m(m\dot{\underline{x}} - \underline{p})\\
\underline{Q} &= \frac{\partial\tilde{L}}{\partial\underline{\dot{p}}} = \dot{\underline{p}} - q\underline{E}_0,
\end{align}

thus

\begin{align}
\tilde{H} &= \underline{P}\underline{\dot{x}} + \underline{Q}\underline{\dot{p}} - \tilde{L} = \nonumber\\
&= \underline{P}\left(\frac{\underline{P}}{\lambda m^2} + \frac{\underline{p}}{m}\right) + \underline{Q}\left(\underline{Q} + q\underline{E}_0\right) - \frac{\underline{Q}^2}{2} - \frac{\underline{P}^2}{2\lambda m^2} = \nonumber\\
&= \frac{\underline{Q}^2}{2} + \frac{\underline{P}^2}{2\lambda m^2} + q\underline{E}_0\underline{Q} + \frac{\underline{p}}{m}\underline{P}
\end{align}

Now we use the fact that

\begin{align}
\frac{\partial\tilde{S}}{\partial t} &= -\tilde{H}\nonumber\\
\frac{\partial\tilde{S}}{\partial\underline{x}} &= \underline{P}\nonumber\\
\frac{\partial\tilde{S}}{\partial\underline{p}} &= \underline{Q}\nonumber
\end{align}

thus - using the notation $K = K'\epsilon_0/q^2$ - for the derivatives of $\tilde{f}$ we obtain

\begin{align}
\frac{\partial \tilde{f}}{\partial t} &= K\tilde{H}\tilde{f}\nonumber\\
\frac{\partial \tilde{f}}{\partial\underline{x}} &= -K\underline{P}\tilde{f}\nonumber\\
\frac{\partial \tilde{f}}{\partial\underline{p}} &= -K\underline{Q}\tilde{f}\nonumber
\end{align}

obtaining

\begin{eqnarray}
\frac{1}{K}\frac{\partial \tilde{f}}{\partial t} = \frac{1}{2K^2}\frac{\partial^2 \tilde{f}}{\partial\underline{p}^2} + \frac{1}{2\lambda K^2 m^2}\frac{\partial^2 \tilde{f}}{\partial\underline{x}^2} - \frac{q\underline{E}_0}{K}\frac{\partial \tilde{f}}{\partial\underline{p}} - \frac{\underline{p}}{Km}\frac{\partial \tilde{f}}{\partial\underline{x}}.
\end{eqnarray}

Simplification with $K$ yields the final form of our differential equation.

\begin{eqnarray}
\frac{\partial \tilde{f}}{\partial t} = \frac{1}{2K}\frac{\partial^2 \tilde{f}}{\partial\underline{p}^2} + \frac{1}{2\lambda Km^2}\frac{\partial^2 \tilde{f}}{\partial\underline{x}^2} - q\underline{E}_0\frac{\partial \tilde{f}}{\partial\underline{p}} - \frac{\underline{p}}{m}\frac{\partial \tilde{f}}{\partial\underline{x}}.\label{diff_final}
\end{eqnarray}

Let us note some interesting properties of this differential equation.

\begin{itemize}

\item The differential equation depends on the phase space coordinates rather then the coordinates of the configuration space.
\item The equation describes a convection--diffusion equation. If there are no external fields present (that is $\underline{E}_0 = 0$) there will still be a convection term left 
since that depends on the variable $\underline{p}$. This means that this equation describes a probability "flow" in phase space where the "velocity" of this "flow" describes 
how the probabilities change.
\item Before we continue let us check the dimensions of the quantities in (\ref{diff_final}). From (\ref{path_density}) the dimension of $K'$ is $m^3/(Js)$ in SI-units (volume per 
action) thus the dimension of $K$ will be $\frac{m^3}{Js}\frac{C^2}{Nm^2}\frac{1}{C^2} = \frac{m}{NmsN}=1/(N^2s)=s^4/(s kg^2 m^2 ) = s^3/(kg^2 m^2 ).$ From (\ref{path_density_mod}) 
the dimension of $\lambda$ will be $1/s^2.$ Now if look at the terms in (\ref{diff_final}) we find that all terms will have dimension $1/s.$ 
\item Equation (\ref{diff_final}) can be rewritten in a more compact form if we introduce generalized coordinates $\underline{z} = (\underline{x},\underline{p})$, that is the 
first three components of $\underline{z}$ are the coordinates, the second are the momenta. If we introduce the matrix

\begin{eqnarray}
\mathbf{D} = \frac{1}{K}
\begin{pmatrix}
\mathbf{1} & \mathbf{0}\\
\mathbf{0} & \frac{1}{\lambda m^2}\mathbf{1}
\end{pmatrix}
\end{eqnarray}

(here each corner is a 3x3 block-diagonal matrix) we can rewrite the second derivative terms as $\frac{1}{2}D_{ij}\frac{\partial^2f}{\partial z_i\partial z_j}$ (all indices go from 
1 to 6). In a similar fashion the first derivative terms can be written as $-V_i\frac{\partial f}{\partial z_i}$ where 

$$\mathbf{V} = \left(\frac{\mathbf{p}}{m}, q\mathbf{E}_0\right).$$

Furthermore if $$\underline{E} = -\nabla_{\underline{x}}\Phi$$ - that is the external electric field is the gradient of a potential - then 

$$V_i = \varepsilon_{ij}\frac{\partial U}{\partial z_i}$$

where $$U=\frac{p^2}{2m} + \Phi$$ is the energy of the system and 

\begin{eqnarray}
\mathbf{\varepsilon} = 
\begin{pmatrix}
\mathbf{0} & -\mathbf{1}\\
\mathbf{1} & \mathbf{0}
\end{pmatrix}
\end{eqnarray}

Putting these together the differential equation will be in the form

\begin{eqnarray}
\frac{\partial \tilde{f}}{\partial t} = \frac{1}{2}D_{ij}\frac{\partial^2\tilde{f}}{\partial z_i\partial z_j} - \varepsilon_{ij}\frac{\partial U}{\partial z_i}\frac{\partial \tilde{f}}{\partial z_i}
\end{eqnarray}

\end{itemize}

\section{Special solutions}

From now we will use the notation $\tilde{f}\to f.$\\

Consider the static problem in two phase space dimensions:

\begin{eqnarray}
0 = \frac{1}{2K}\frac{\partial^2 f}{\partial p^2} + \frac{1}{2Km^2\lambda}\frac{\partial^2 f}{\partial x^2} - \frac{p}{m}\frac{\partial f}{\partial x} - qE\frac{\partial f}{\partial p},\label{basiceq2d}
\end{eqnarray}

where $f = f(x,p)$ is the probability density (thus it is required to be positive everywhere and its integral should be finite), $E = E(x)$ is the electric field, 
$m$ is the mass, $q$ is the elementary charge, $\lambda > 0$ is a parameter and $K$ is a positive constant.\\

In the following some of the most basic problems in quantum mechanics will be discussed and it will be shown that - at least qualitatively - this model perfectly describes 
the basic phenomena occurring in theses systems. 

\subsection{Case 1: Free particle}

This is the case where $E = 0$ so our differential equation will be of the form

\begin{eqnarray}
0 = \frac{1}{2K}\frac{\partial^2 f}{\partial p^2} + \frac{1}{2Km^2\lambda}\frac{\partial^2 f}{\partial x^2} - \frac{p}{m}\frac{\partial f}{\partial x}\label{free_eq}
\end{eqnarray}

Let us search for solutions of the form $f_\beta(x,p) = e^{-i\beta x}g_\beta(p)$ where $\beta$ is real (this is a necessary condition for the solution to be bounded). Substituting 
this into (\ref{basiceq2d}) we obtain the following differential equation for $g_\beta(p):$

\begin{eqnarray}
0 = \frac{d^2 g_\beta}{dp^2} - \frac{2K\beta}{m}\left(\frac{\beta}{2Km\lambda} - ip\right)g_\beta.
\end{eqnarray}

Introducing the new variable $$\tilde{p} :=\left(\frac{2K\beta}{m}\right)^3\left(\frac{\beta}{2Km\lambda} - ip\right)$$ the differential equation will take the form

\begin{eqnarray}
\frac{d^2 g_\beta}{d\tilde{p}^2} - \tilde{p}g_\beta = 0
\end{eqnarray}

which is the Airy differential equation with solutions 

\begin{eqnarray}
g_\beta(\tilde{p}) = \frac{1}{2i\pi}\int_{\Gamma} dt\exp\left(\frac{it^3}{3} + it\tilde{p}\right).\label{Airy_def}
\end{eqnarray}

with $\Gamma$ being a contour in the complex plane. Though it seems like there is only one solution, it can be shown (e.g. in \cite{airy_appendix_E}) that the choices in the 
contour will give us two independent solutions. In our case the necessary condition is that the integral of the solution should be bounded (otherwise it would not be a probability 
density function), which means that we have to chose $\Gamma$ so that the integral of $g_\beta(\tilde{p})$ is finite for every $\beta$. To find this contour first let us 
define

\begin{align}
N(\beta) &= -\left(\frac{2K\beta}{m}\right)^{1/3}\label{N_beta_def}\\
p_0(\beta) &= -\frac{\beta}{2Km\lambda}.\label{p_0_beta_def}
\end{align}

thus 

$$g_\beta(\tilde{p}) = g_\beta(N(\beta)(p_0(\beta) + ip)).$$

It is easy to show that one such a contour is the complex line. Indeed, if one choses the complex line, that is changes the integration variable $t$ to $it$ then 

\begin{align}
\int_{-\infty}^\infty g_\beta(N(p_0 + ip))dp &= \frac{1}{2\pi}\int_{-\infty}^\infty dp\int_{-\infty}^\infty dt\exp\left(\frac{t^3}{3} - t(N(p_0 + ip))\right) = \nonumber\\
&= \frac{1}{2\pi}\int_{-\infty}^\infty dp \int_{-\infty}^{\infty} dt \exp\left(\frac{t^3}{3} - tNp_0 - itNp\right) = \frac{1}{N},\label{Airy_norm}
\end{align}

where in the last step we used the fact that $(1/(2\pi))\int_{-\infty}^\infty dpe^{ipt} = \delta(t)$. Checking other contours is not necessary for the following reason. Since 
two solutions of the Airy-functions are independent if and only if they cross different sectors defined by Stokes-lines (see in \cite{airy_three_sector}) it follows that (since 
there are only three regions) another solution would not follow the complex line in either $-\infty$ or $\infty$ which would mean that the integral of $g_\beta(N(p_0 + ip))$ 
would be infinite.\\

In order to distinguish $g_\beta(z)$ from the usual definitions of the Airy-function we will denote it with $\tilde{Ai}(z).$ Thus the complete solution of the differential 
equation (\ref{free_eq}) will be 

\begin{eqnarray}
f(x,p) = \int_{-\infty}^\infty C(\beta)e^{-i\beta x}\tilde{Ai}(N(\beta)(p_0(\beta) + ip)) d\beta\label{free_eq_solution_before_norm}
\end{eqnarray}

where $\tilde{Ai}(z)$ is defined via (\ref{Airy_def}), the contour $\Gamma$ being the complex line, that is

\begin{eqnarray}
\tilde{Ai}(z) = \frac{1}{2\pi}\int_{-\infty}^\infty \exp\left(\frac{t^3}{3} - tz\right)dt,\label{Airy_tilde_def}
\end{eqnarray}

$N(\beta)$ and $p_0(\beta)$ are defined via (\ref{N_beta_def}) and (\ref{p_0_beta_def}) respectively and $C(\beta)$ is a (currently) arbitrary complex function.\\

The above expression is not normalized so we have to calculate the norm first:

\begin{align}
\tilde{N} &= \int_{-\infty}^\infty dx \int_{-\infty}^\infty dp \int_{-\infty}^\infty d\beta C(\beta)e^{-i\beta x}\tilde{Ai}(N(\beta)(p_0(\beta) + ip)) = \nonumber\\
&= 2\pi\int_{-\infty}^\infty d\beta \frac{\delta(\beta)C(\beta)}{N(\beta)} = \lim_{\beta\to 0}\frac{C(\beta)}{N(\beta)} = -\left(\frac{m}{2K}\right)^{1/3}\lim_{\beta\to 0}\frac{C(\beta)}{\beta^{1/3}} \label{f_x_p_norm}
\end{align}

thus the final expression for $f(x,p)$ will be 

\begin{eqnarray}
f(x,p) = \frac{1}{\tilde{N}}\int_{-\infty}^\infty C(\beta)e^{-i\beta x}\tilde{Ai}(N(\beta)(p_0(\beta) + ip)) d\beta\label{free_eq_solution}
\end{eqnarray}

where $\tilde{N}$ is defined via (\ref{f_x_p_norm}).

\subsubsection{Properties}

Let us examine the properties of this solution.

\begin{itemize}
\item $\beta$ has to be real in order to have a bounded solution.
\item It is easy to show that the solution is real if $C(\beta)^{*} = C(-\beta).$

\item Let us calculate the marginal distributions.\\
First let us calculate $f_x(x),$ the marginal distribution of the coordinate. By definition 

\begin{eqnarray}
f_x(x) = \int_{-\infty}^\infty f(x,p)dp = \int_{-\infty}^\infty \frac{C(\beta)}{\tilde{N}N(\beta)}e^{-i\beta x} d\beta\label{marginal_x}
\end{eqnarray}

where we used the result of equation (\ref{Airy_norm}). Thus it turns out that $C(\beta)/(\tilde{N}N(\beta))$ is the characteristic function of the marginal distribution $f_x$!\\
Calculating the marginal distribution $f_p(p)$ is simpler:

\begin{eqnarray}
f_p(p) = \int_{-\infty}^\infty f(x,p)dx = \frac{2\pi}{\tilde{N}}\int_{-\infty}^\infty C(\beta)\delta(\beta)\tilde{Ai}(N(\beta)(p_0(\beta) + ip))d\beta
\end{eqnarray}

Imposing the condition for $C(\beta)$ that 

$$\lim_{\beta\to 0}C(\beta)\tilde{Ai}(N(\beta)(p_0(\beta) + ip)) := \frac{\tilde{N}\tilde{C}(p)}{2\pi} = finite$$

we obtain the formula for $f_p(p)$:

\begin{eqnarray}
f_p(p) = \tilde{C}(p).\label{marginal_p}
\end{eqnarray}

This formula can only be made explicit if we know the function $C(\beta)$ thus the limit $\lim_{\beta\to 0}C(\beta)\tilde{Ai}(N(\beta)(p_0(\beta) + ip))$ can be calculated. 

\item With the help of the marginal distributions we can give the formulas for the conditional distributions.\\
First let $x = X.$ Then the conditional distribution function $f(x,p|x = X)$ will be

\begin{align}
f(x,p|x = X) &= \frac{f(X,p)}{f_x(X)} = \frac{\int_{-\infty}^\infty C(\beta)e^{-i\beta X}\tilde{Ai}(N(\beta)(p_0(\beta) + ip)) d\beta}{\int_{-\infty}^\infty \frac{C(\beta)}{N(\beta)}e^{-i\beta X} d\beta} = \nonumber\\
&= \int_{-\infty}^\infty F(\beta, X)\tilde{Ai}(N(\beta)(p_0(\beta) + ip)) d\beta\label{conditional_dist_X}
\end{align}

where we have defined

\begin{eqnarray}
F(\beta, X) = \frac{C(\beta)e^{-i\beta X}}{\int_{-\infty}^\infty \frac{C(\beta)}{N(\beta)}e^{-i\beta X} d\beta}.\label{Big_F_def_cont}
\end{eqnarray}

Now let $p = P.$ In this case the conditional distribution function $f(x,p|p = P)$ will be

\begin{eqnarray}
f(x,p|p = P) &= \frac{f(x,P)}{f_p(P)} = \int_{-\infty}^\infty \frac{C(\beta)}{\tilde{C}(P)}e^{-i\beta x}\tilde{Ai}(N(\beta)(p_0(\beta) + iP)) d\beta\label{conditional_dist_P}
\end{eqnarray}

Note that the normalization $\tilde{N}$ disappears from the conditional distributions (since it appears in both the denominator and the nominator of the definition).

\item From (\ref{marginal_x}) and (\ref{marginal_p}) it follows that the moments of $x$ are calculated by the derivation of $C(\beta)/N(\beta)$, evaluated by zero and multiplied 
by $(-i)^n$ where $n$ is the order of the moment. Since the expression $f_p(p)$ is only implicit at the moment it is meaningless to talk about its moments.\\
However the moments of the conditional probabilities exist for both $x$ and $p$. For the former 

\begin{align}
<x|p = P> &= \int_{-\infty}^\infty x f(x,p|p = P) dx = \nonumber\\
&= \int_{-\infty}^\infty \frac{C(\beta)}{\tilde{C}(P)}\left(\frac{1}{-i}\frac{d}{d\beta}\int_{-\infty}^\infty e^{-i\beta x}dx\right) \tilde{Ai}(N(\beta)(p_0(\beta) + iP)) d\beta = \nonumber\\
&= (-i)\frac{d}{d\beta}\left[ \frac{C(\beta)}{\tilde{C}(P)}\tilde{Ai}(N(\beta)(p_0(\beta) + iP))\right]\bigg|_{\beta = 0}
\end{align}

In a similar fashion

\begin{align}
<x^n|p = P> = (-i)^n\frac{d^n}{d\beta^n}\left[ \frac{C(\beta)}{\tilde{C}(P)}\tilde{Ai}(N(\beta)(p_0(\beta) + iP))\right]\bigg|_{\beta = 0}
\end{align}

To calculate the moments with respect to $p$ we need to calculate the integrals $\int p\tilde{Ai}(N(p_0 + ip))dp$ and $\int p^2\tilde{Ai}(N(p_0 + ip))dp.$ These can 
be obtained in a similar fashion to the case when we calculated $\int \tilde{Ai}(N(p_0 + ip))dp$ (the multiplication with $N$ is due to the fact that the calculation of the 
moment involves a division with respect to the norm, which is now $1/N$ (see (\ref{Airy_norm}))).

\begin{align}
N\int_{-\infty}^\infty &pAi(N(p_0 + ip))dp = \frac{N}{2\pi}\int_{-\infty}^\infty dp \int_{-\infty}^{\infty} dt p\exp\left(\frac{t^3}{3} - tN(p_0 + ip)\right) = \nonumber\\
&= \int_{-\infty}^{\infty} dt \exp\left(\frac{t^3}{3} - tNp_0 \right)\frac{N}{2\pi}\int_{-\infty}^\infty p\exp(-itNp) dp  = \nonumber\\
&= \int_{-\infty}^{\infty} dt \exp\left(\frac{t^3}{3N^3} - tp_0\right) \frac{d\delta(t)}{dt}  = \nonumber\\
&= p_0\label{first_moment}\\
N\int_{-\infty}^\infty &p^2Ai(N(p_0 + ip))dp = \frac{N}{2\pi}\int_{-\infty}^\infty dp \int_{-\infty}^{\infty} dt p^2\exp\left(\frac{t^3}{3} - tN(p_0 + ip)\right) = \nonumber\\
&= \int_{-\infty}^{\infty} dt \exp\left(\frac{t^3}{3} - tNp_0\right) \frac{N}{2\pi}\int_{-\infty}^\infty p^2\exp(-itNp) dp  = \nonumber\\
&= \int_{-\infty}^{\infty} dt \exp\left(\frac{t^3}{3N^3} - tp_0\right) \frac{d^2\delta(t)}{dt^2}  = \nonumber\\
&= p^2_0\label{second_moment}
\end{align}

Using the formulas (\ref{first_moment}) and (\ref{second_moment}) the first and second conditional moments of $f(x,p|x = X)$ will be

\begin{align}
<p|x = X> = \int_{-\infty}^\infty p f(x,p|x = X) dp = \int_{-\infty}^\infty \frac{F(\beta, X)}{N(\beta)}p_0(\beta)d\beta\label{first_cond_moment}\\
<p^2|x = X> = \int_{-\infty}^\infty p^2 f(x,p|x = X) dp = \int_{-\infty}^\infty \frac{F(\beta, X)}{N(\beta)}p^2_0(\beta)d\beta\label{second_cond_moment}
\end{align}

thus the first and second moment of the conditional probability density $f(x,p|x = X)$ can be expressed with $p_0(\beta).$ 

\item It is not trivial as to under what conditions (what choice of $C(\beta)$ functions) would (\ref{free_eq_solution}) be everywhere non-negative. Though there are theorems for a 
Fourier-integral to be positive (e.g. Bochner's theorem) the arising conditions for $C(\beta)$ can not be solved explicitly in this case. Thus the question is open whether there 
are not only necessary but sufficient conditions for the coefficients so that the solution can be a probability density function.

\end{itemize}

\subsection{Case 2: Infinite potential well}

The infinite potential well can be modeled with $E(x) = E_0(\delta(x - a) - \delta(x + a))$ where $E_0 = const.$ First we will start the calculation as if $E_0$ were finite, 
and after that we will take the limit $E_0\to\infty.$ \\
Let us substitute the expression of $E(x)$ into \ref{basiceq2d} and integrate the whole equation with respect to x from $-\infty$ to $\infty$. We get

\begin{eqnarray}
0 = \frac{1}{2K}\frac{d^2 f_p(p)}{dp^2} - qE_0\left[\frac{\partial f}{\partial p}\bigg|_{x = a} - \frac{\partial f}{\partial p}\bigg|_{x = -a}\right],
\end{eqnarray}

where $f_p(p)=\int_{-\infty}^{\infty}f(x,p)dx$ is the marginal distribution for $p$ and we used the fact that for all $p$ the derivatives of $f(x,p)$ will be zero at infinity. 
Rearranging this equation yields 

\begin{eqnarray}
\frac{1}{2KE_0}\frac{d^2 f_p(p)}{dp^2} = q\left[\frac{\partial f}{\partial p}\bigg|_{x = a} - \frac{\partial f}{\partial p}\bigg|_{x = -a}\right],
\end{eqnarray}

thus taking the limit $E_0\to\infty$ we obtain the following boundary condition on f:

\begin{eqnarray}
0 = \frac{\partial f}{\partial p}\bigg|_{x = a} - \frac{\partial f}{\partial p}\bigg|_{x = -a}\label{boundary}.
\end{eqnarray}

Since our equation is local, the term containing the electric field will not contribute to the equation if $x$ is does not equal $a$ or $-a$ leaving us with the case of the 
free particle solution (\ref{free_eq_solution}):

\begin{eqnarray}
f(x,p) = \frac{1}{\tilde{N}}\int_{-\infty}^\infty C(\beta)e^{-i\beta x}\tilde{Ai}(N(\beta)(p_0(\beta) + ip)) d\beta.
\end{eqnarray}

Substituting the boundary conditions into this equation we have 

\begin{eqnarray}
\frac{1}{\tilde{N}}\int_{-\infty}^\infty C(\beta)\left(e^{-i\beta a} - e^{i\beta a}\right)\frac{d}{dp}\tilde{Ai}(N(\beta)(p_0(\beta) + ip)) d\beta = 0
\end{eqnarray}

thus - since the boundary conditions have to be true for all $p$ - we obtain the following condition for $\beta$:

\begin{eqnarray}
\beta = \frac{n\pi}{a},
\end{eqnarray}

where $n$ is an integer. This means that in this case the only valid solutions are those where $C(\beta)$ is of the form

\begin{eqnarray}
C(\beta) = \sum_{n = -\infty}^\infty C_n\delta(\beta - \frac{n\pi}{a})\label{C_beta_discrete}
\end{eqnarray}

thus we may rewrite the form of our solution as 

\begin{eqnarray}
f(x,p) = \frac{1}{\tilde{N}}\sum_{n=-\infty}^n C_n e^{-in\pi\frac{x}{a}}\tilde{Ai}\left(N\left(\frac{n\pi}{a}\right)(p_0\left(\frac{n\pi}{a}\right) + ip)\right).
\end{eqnarray}

\subsubsection{Properties}

\begin{itemize}

\item We have seen that the solution of the free particle is real if $C(\beta)^{*} = C(-\beta)$ which translates to this case as $C_n^{*} = C_{-n}.$

\item The marginal distributions and conditional distributions can be derived easily if we substitute the expression (\ref{C_beta_discrete}) into the appropriate formula for 
the free particle.

\item Specifically let us calculate the expectation value of the kinetic energy $p^2/(2m)$ with the condition that $x = X$ for some fixed $X$. First let us define 

\begin{eqnarray}
F_j(X) = \frac{C_je^{-i\frac{jX\pi}{a} }}{\sum_{n = -\infty}^\infty \frac{C_n}{N_n}e^{-i\frac{nX\pi}{a} }}.
\end{eqnarray}

which is the potential well version of formula (\ref{Big_F_def_cont}) (here we have used the notation $N_n = N(\frac{n\pi}{a})$). Now for the conditional expectation value for 
the second moment of the momentum we will obtain a similar formula to (\ref{second_cond_moment}), so the conditional expectation value of the kinetic energy will be 

\begin{align}
<E_{kin}|x = X> &= \frac{<p^2|x = X>}{2m} = \frac{1}{2m}\sum_{n = -\infty}^\infty \frac{F_n(X)}{N_n}p^2_0\left(\frac{n\pi}{a}\right) = \nonumber\\
&= \frac{1}{2m}\sum_{n = -\infty}^\infty \frac{F_n(X)}{N_n}\frac{n^2\pi^2}{(2Km\lambda)^2a^2}.
\end{align}

where in the last step we have used the definition (\ref{p_0_beta_def}) of $p_0.$\\

Defining 

\begin{eqnarray}
\tilde{F}_j(X) := \frac{F_j(X)}{N_j} = \frac{C_je^{-i\frac{jX\pi}{a} }}{\sum_{n = -\infty}^\infty C_n\left(\frac{j}{n}\right)^{1/3}e^{-i\frac{nX\pi}{a} }}
\end{eqnarray}

(here we used the definition (\ref{N_beta_def}) of $N(\beta)$) we may rewrite this expression as 

\begin{eqnarray}
<E_{kin}|x = X> = \sum_{n = -\infty}^\infty \tilde{F}_n(X)\frac{1}{2m}\frac{1}{(2Km\lambda)^2}\frac{n^2\pi^2}{a^2}\label{pot_well_energy_random}
\end{eqnarray}

The terms after the multiplication factor $\tilde{F}_n(X)$ are very similar to the energy levels obtained in quantum mechanics. There the energy eigenvalues in the infinite 
potential well are 

\begin{eqnarray}
E^n_{QM} = \frac{\bar{h}^2}{2m}\frac{n^2\pi^2}{a^2}\label{pot_well_energy_QM}
\end{eqnarray}

so if we identify $\bar{h}$ with $1/(2Km\lambda)$ the same expression is obtained. The dimensions are the same since from the properties of the differential equation for $f(x,p)$ 
we find that 

\begin{eqnarray}
dim\left(\frac{1}{Km\lambda}\right) = \frac{kg^2m^2}{s^3}\frac{1}{kg}s^2 = \frac{kg m^2}{s}
\end{eqnarray}

\end{itemize}

\textbf{Interpretation} How could we interpret these results?\\
At first the fact that we obtained the same result as in quantum mechanics if we identified $\bar{h}$ with $1/(2Km\lambda)$ seems to be troubling since the latter term 
contains the mass of the particle. However it also contains the constant $\lambda$ which may depend on the mass of the particle. Specifically if we define 
$\tilde{\lambda} = m\lambda$ in the starting Lagrangian then the final expressions will change only in the constant factors, so then $\bar{h}$ will equal $1/(2K\tilde{\lambda}).$\\
Also it is interesting that we have obtained a weighed sum instead of specific values. This however is not a surprise. Recall that our expression for the distribution density 
resembles a path integral. Thus every quantity which is calculated via that density function can be interpreted as a sum of values for a specific path. Specifically if we look 
at the expression $f(x,p|x = X)$ this takes into account all possible paths that go through $X$ and contains information about the "weight" of that specific path for each $p.$ 
Thus if the expectation value of $p^2$ is calculated then this quantity can be interpreted as the weighed average for all the momenta for each path.

\subsection{Case 3: Harmonic oscillator}

Let us perform a Fourier-transformation to (\ref{basiceq2d}): multiply the equation with $\exp(-ipk - ixl)$ and define $F(l,k)=\frac{1}{2\pi}\int f(x,p)\exp(-ipk - ixl)dpdx.$ 

Using the following properties

\begin{align}
\mathcal{F}\left(\frac{\partial f}{\partial x}\right) = (il)F\nonumber\\
\mathcal{F}(pf) = i\frac{\partial F}{\partial k}\nonumber\\
\mathcal{F}(E(x)f) = \frac{\tilde{E}(l)*F(l,k)}{\sqrt{2\pi}}\nonumber
\end{align}

($\mathcal{F}$ denotes the Fourier-transform, $\tilde{E}(l)$ is the Fourier-transform of $E(x)$ and $\tilde{E}*F$ is the convolution of the two function) we obtain 

\begin{eqnarray}
0 = -\frac{k^2}{2K}F - \frac{l^2}{2Km^2\lambda}F + \frac{l}{m}\frac{\partial F}{\partial k} - \frac{ikq\tilde{E}*F}{\sqrt{2\pi}}.
\end{eqnarray}

In the case of the harmonic oscillator $E(x) = -Dx$ where $D$ is some constant so $E(l) = -iD\sqrt{2\pi}d\delta(l)/dl$ thus

\begin{eqnarray}
0 = -\frac{k^2}{2K}F - \frac{l^2}{2Km^2\lambda}F + \frac{l}{m}\frac{\partial F}{\partial k} - kDq\frac{\partial F}{\partial l}.
\end{eqnarray}

This is a first order partial differential equation for which the characteristic equations are

\begin{align}
\frac{1}{F}\frac{dF}{dk} &= -\frac{m}{l}\left(\frac{k^2}{2K} + \frac{l^2}{2Km^2\lambda}\right)\\
\frac{1}{F}\frac{dF}{dl} &= \frac{1}{kDq}\left(\frac{k^2}{2K} + \frac{l^2}{2Km^2\lambda}\right)\\
\frac{dk}{dl} &= -\frac{l}{kmDq}
\end{align}

From the third equation we can obtain the integral curves:

$$\frac{1}{2}(mDqk^2 + l^2) - A = 0.$$ 

If $Dq > 0$ this formula suggests to look for a substitution of the form

\begin{align}
k &= r\sin\phi\nonumber\\
l &= \sqrt{mDq}r\cos\phi\nonumber
\end{align}

After the substitution we obtain

\begin{eqnarray}
0 = -\frac{r^2}{2K}\left(\sin^2\phi + \frac{Dq}{m\lambda}\cos^2\phi\right)F + \sqrt{\frac{Dq}{m}}\frac{\partial F}{\partial\phi}\label{harmonic_de_polar}
\end{eqnarray}

thus 

\begin{align}
F(r,\phi) &= \tilde{B}(r)\exp\left[\frac{r^2}{4K}\sqrt{\frac{m}{Dq}}\left(\phi - \sin\phi\cos\phi + \frac{Dq}{m\lambda}(\phi + \sin\phi\cos\phi)\right)\right] = \nonumber\\
&= \tilde{B}(r)\exp\left[\frac{r^2}{4K}\sqrt{\frac{m}{Dq}}\left(\phi\left(\frac{Dq}{m\lambda} + 1\right) + \frac{\sin(2\phi)}{2}\left(\frac{Dq}{m\lambda} - 1\right)\right)\right] = \nonumber\\
&= \exp\left[\alpha r^2\left((\phi + B(r))(\beta + 1) + \frac{\sin(2\phi)}{2}(\beta - 1)\right)\right]\label{harmonic_char}
\end{align}

where in the last step we introduced an alternative integration constant $B(r)$ so that it can be in the exponent of the expression (this integration constant has to depend on $r$ 
since (\ref{harmonic_de_polar}) contains a derivative only with respect to $\phi$) and defined $\alpha = \frac{1}{4K}\sqrt{\frac{m}{Dq}},\ \beta = Dq/(m\lambda)$.\\

Before we continue to check the properties of the result let us write $F$ in terms of $k$ and $l:$

\begin{align}
F(l,k) &= \tilde{B}\left(k^2 + \frac{l^2}{mDq}\right)\exp\left[\alpha\left(k^2 + \frac{l^2}{mDq}\right)(\beta + 1)\left(Arctg\frac{k\sqrt{mDq}}{l}\right)\right. + \nonumber\\
&+ \left.\alpha(\beta - 1)\frac{kl}{\sqrt{mDq}}\right].\label{harmonic_char2}
\end{align}

\subsubsection{Properties}

\begin{itemize}

\item Since $F(l,k)$ is the Fourier-transform of $f(x,p)$ then this is the characteristic function of the distribution density. From this follows that various calculations involve 
evaluating $F(l,k)$ and its derivatives at those points where either $k$ or $l$ (or both) are zero. This is quite difficult if look at (\ref{harmonic_char2}) since the expression 
$Arctg\frac{k\sqrt{mDq}}{l}$ does not have a limit at $l = 0,\ k = 0.$ Because of this we will use the expression (\ref{harmonic_char}) instead since then the value of $F(r, \phi)$ 
can be easily calculated at the required points.\\
Indeed if we look at what values correspond to the zeros of $l$ and $k$ we obtain the following:

\begin{align}
l &= 0 \wedge k\neq 0 \rightarrow \phi = \left(n + \frac{1}{2}\right)\pi \wedge r\neq 0\nonumber\\
l &\neq 0 \wedge k = 0 \rightarrow \phi = n\pi \wedge r\neq 0\nonumber\\
l &= 0 \wedge k = 0 \rightarrow r = 0\nonumber\\
\end{align}

We see two things.\\
First because of the undefined limit of the $Arctg$ function multiple values of $\phi$ represent a zero if only one of the variables are zero. This can be interpreted in the 
following way. If we look at (\ref{harmonic_char}) we see that the expression is not periodic because the $\alpha r^2\phi(\beta + 1)$ term in the exponent. This means that in 
order to obtain all values of the expression we need to extend this function analytically to the $\phi\in [-\infty,\ \infty]$ interval. This can be done in a straightforward 
manner since the $\alpha r^2\phi(\beta + 1)$ term is linear in $\phi$ while the other terms are either constants or periodic.\\
Second the cases where only one variable is zero or both are separated in the sense that the latter imposes a condition only on the variable $r$ while the former only on $\phi.$

\item The condition $F(l = 0, k = 0) = 1$ will imply that $F(r = 0, \phi) = 1$ which means that 

$$\lim_{r \to 0} r^2 B(r) = 0$$ 

thus $B(r)$ can have at most a first order pole in $r = 0.$

\item It is easy to show that the characteristic functions of the marginal distributions can be obtained by setting one of the variables zero. This means that 

\begin{align}
F_x(r) &:= F(r, \phi = n\pi) = \exp\left[\alpha r^2(n\pi + B(r))(\beta + 1)\right]\\
F_p(r) &:= F\left(r, \phi = \left(n + \frac{1}{2}\right)\pi\right) = \exp\left[\alpha r^2\left(\left(n + \frac{1}{2}\right)\pi + B(r)\right)(\beta + 1)\right]
\end{align}

will be the characteristic functions of $f_x(x)$ and $f_p(p)$ respectively.

\item The first and second moments are calculated from the first and second partial derivatives of $F(l,k)$ respectively so we need to express these in terms of $r$ and $\phi.$ 
The first derivatives of $r$ and $\phi$ with respect to $k$ and $l$ are the following: 

\begin{align}
\frac{\partial r}{\partial k} &= \sin\phi\nonumber\\
\frac{\partial\phi}{\partial k} &= \frac{\frac{\sqrt{mDq}}{l}}{1 + \frac{k^2mDq}{l^2}} = \frac{\cos\phi}{r}\nonumber\\
\frac{\partial r}{\partial l} &= \frac{\cos\phi}{\sqrt{mDq}}\nonumber\\
\frac{\partial\phi}{\partial l} &= \frac{-\frac{k\sqrt{mDq}}{l^2}}{1 + \frac{k^2mDq}{l^2}} = -\frac{\sin\phi}{r\sqrt{mDq}}\nonumber
\end{align}

Then for the first derivatives of $F$ we yield the following expressions:

\begin{align}
\frac{\partial F}{\partial k} &= \frac{\partial r}{\partial k}\left[2\alpha r\left((\phi + B(r))(\beta + 1) + \frac{\sin(2\phi)}{2}(\beta - 1)\right) + \alpha r^2 B'(r)(\beta + 1)\right]F(r,\phi) + \nonumber\\
&+ \frac{\partial\phi}{\partial k}\left[\alpha r^2\left((\beta + 1) + \cos(2\phi)(\beta - 1)\right)\right]F(r,\phi) = \nonumber\\
&= \sin\phi\left[2\alpha r\left((\phi + B(r))(\beta + 1) + \frac{\sin(2\phi)}{2}(\beta - 1)\right) + \alpha r^2 B'(r)(\beta + 1)\right]F(r,\phi) + \nonumber\\
&+ \cos\phi\alpha r\left[(\beta + 1) + \cos(2\phi)(\beta - 1)\right]F(r,\phi)\label{harmonic_partial_k}\\
\frac{\partial F}{\partial l} &= \frac{\partial r}{\partial l}\left[2\alpha r\left((\phi + B(r))(\beta + 1) + \frac{\sin(2\phi)}{2}(\beta - 1)\right) + \alpha r^2 B'(r)(\beta + 1)\right]F(r,\phi) + \nonumber\\
&+ \frac{\partial\phi}{\partial l}\left[\alpha r^2\left((\beta + 1) + \cos(2\phi)(\beta - 1)\right)\right]F(r,\phi) = \nonumber\\
&= \frac{\cos\phi}{\sqrt{mDq}}\left[2\alpha r\left((\phi + B(r))(\beta + 1) + \frac{\sin(2\phi)}{2}(\beta - 1)\right) + \alpha r^2 B'(r)(\beta + 1)\right]F(r,\phi) - \nonumber\\
&- \frac{\sin\phi\alpha r}{\sqrt{mDq}}\left[(\beta + 1) + \cos(2\phi)(\beta - 1)\right]F(r,\phi)\label{harmonic_partial_l}
\end{align}

Notice that if we introduce the quantities

\begin{align}
A_r(r,\phi) &= 2\alpha r\left((\phi + B(r))(\beta + 1) + \frac{\sin(2\phi)}{2}(\beta - 1)\right) + \alpha r^2 B'(r)(\beta + 1)\\
A_\phi(r,\phi) &= \alpha r\left[(\beta + 1) + \cos(2\phi)(\beta - 1)\right]
\end{align}

we can rewrite the first derivatives in the following form:

\begin{align}
\frac{\partial F}{\partial k} &= (\sin\phi A_r + \cos\phi A_\phi)F\label{harmonic_partial_k_simple}\\
\frac{\partial F}{\partial l} &= \frac{1}{\sqrt{mDq}}(\cos\phi A_r - \sin\phi A_\phi)F\label{harmonic_partial_l_simple}
\end{align}

\item We see immediately from expressions (\ref{harmonic_partial_k}) and (\ref{harmonic_partial_l}) that the existence of the expectation value depends on the behavior of $B'(r)$ 
at the origin. Indeed the expectation value is proportional to 

$$\frac{\partial F}{\partial k}\bigg|_{k = 0, l = 0}, \frac{\partial F}{\partial k}\bigg|_{k = 0, l = 0}$$

which is expressed in terms with our new variables as calculating (\ref{harmonic_partial_k}) and (\ref{harmonic_partial_l}) at $r = 0.$ This yields

\begin{align}
\frac{\partial F}{\partial k}\bigg|_{r = 0} = \alpha(\beta + 1)\sin\phi\lim_{r\to 0}[2rB(r) + r^2 B'(r)]\\
\frac{\partial F}{\partial l}\bigg|_{r = 0} = \frac{\alpha(\beta + 1)\cos\phi}{\sqrt{mDq}}\lim_{r\to 0}[2rB(r) + r^2 B'(r)].
\end{align}

\item The conditional distributions are hard to calculate since they are the quotient of two functions. However we will show that the respective conditional expectation values are 
proportional to taking the derivative of $F$ with respect to a variable and taking only $k$ or $l$ as zero thus finding the characteristic function of the conditional distributions 
is unnecessary.\\
Let us start from definition of the conditional expectation value of $f(x,p|p = P)$ where $P$ is a fixed value.

\begin{align}
<x|p = P> &= \int_{-\infty}^\infty xf(x,p|p = P)dx = \int_{-\infty}^\infty \frac{xf(x,P)}{f_p(P)}dx = \nonumber\\
&= \int_{-\infty}^\infty dx\int_{-\infty}^\infty dl\int_{-\infty}^\infty dk \frac{xF(l,k)e^{ixl+iPk}}{\sqrt{2\pi}f_p(P)} = \nonumber\\
&= \int_{-\infty}^\infty dx\int_{-\infty}^\infty dl\int_{-\infty}^\infty dk \frac{\frac{\partial F(l,k)}{\partial l}e^{ixl+iPk}}{-i\sqrt{2\pi}f_p(P)} = \nonumber\\
&= \int_{-\infty}^\infty dl\int_{-\infty}^\infty dk \frac{\frac{\partial F(l,k)}{\partial l}\big|_{l = 0}e^{iPk}}{-if_p(P)}
\end{align}

where in the last two steps we have used the formula $xe^{ilx} = \frac{1}{i}\frac{d}{dl}e^{ilx},$ integrated by parts with respect to $l$ and used the fact that 
$\int dx e^{ixl} = \sqrt{2\pi}\delta(l)$ which means that the conditional expectation value of $x$ will be proportional to the quantity

$$\frac{\partial F}{\partial l}\bigg|_{l = 0}.$$

In a similar fashion one can show a similar results to higher order momenta of $x$ (in that case the respected higher partial derivatives will appear) and the conditional 
expectation values for $p.$\\

Using this result the conditional probabilities will be proportional to

\begin{align}
\frac{\partial F}{\partial l}\bigg|_{l = 0} &= \frac{1}{\sqrt{mDq}}(\cos\phi A_r - \sin\phi A_\phi)F|_{\phi = \left(n + \frac{1}{2}\right)\pi} = (-1)^{n+1}\frac{2\alpha r}{\sqrt{mDq}}\\
\frac{\partial F}{\partial k}\bigg|_{k = 0} &= (\sin\phi A_r + \cos\phi A_\phi)F|_{\phi = n\pi} = (-1)^n\alpha r(\beta + 1)
\end{align}

\item With the help of the expressions (\ref{harmonic_partial_k_simple}) and (\ref{harmonic_partial_l_simple}) we can calculate the second derivatives of $F.$ First let us 
define the following quantities:

\begin{align}
A_{rr} &:= \frac{\partial A_r}{\partial r} = 2\alpha\left((\phi + B(r))(\beta + 1) + \frac{\sin(2\phi)}{2}(\beta - 1)\right) + \nonumber\\
&+ 4\alpha r B'(r)(\beta + 1) + \alpha r^2 B''(r)(\beta + 1)\\
A_{r\phi} &:= \frac{\partial A_r}{\partial\phi} = 2\alpha r\left[(\beta + 1) + \cos(2\phi)(\beta - 1)\right]\\
A_{\phi r} &:= \frac{\partial A_\phi}{\partial r} = \alpha\left[(\beta + 1) + \cos(2\phi)(\beta - 1)\right]\\
A_{\phi\phi} &:= \frac{\partial A_\phi}{\partial\phi} = -2\alpha r\sin(2\phi)(\beta - 1)
\end{align}

Now let us calculate the second derivatives:

\begin{align}
\frac{\partial^2F}{\partial k^2} &= \left[\frac{\cos^2\phi}{r}A_r + \sin^2\phi A_{rr} + \frac{\sin\phi\cos\phi}{r} A_{r\phi} - \right.\nonumber\\
&- \left.\frac{\sin\phi\cos\phi}{r}A_\phi + \sin\phi\cos\phi A_{\phi r} + \frac{\cos^2\phi}{r}A_{\phi\phi} + \right.\nonumber\\
&+ \left.(\sin\phi A_r + \cos\phi A_\phi)^2\right]F\label{F_kk}\\
\frac{\partial^2F}{\partial l^2} &= \frac{1}{mDq}\left[\frac{\sin^2\phi}{r} A_r + \cos^2\phi A_{rr} - \frac{\sin\phi\cos\phi}{r} A_{r\phi} + \right.\nonumber\\
&+ \left. \frac{\sin\phi\cos\phi}{r}A_\phi - \sin\phi\cos\phi A_{\phi r} + \frac{\sin^2\phi}{r}A_{\phi\phi} + \right.\nonumber\\
&+ \left.(\cos\phi A_r - \sin\phi A_\phi)^2\right]F\label{F_ll}
\end{align}

\item The interesting quantity to calculate is the expectation value of the energy. In our notations this will be the expectation value of 

$$\frac{p^2}{2m} + \frac{qD}{2}x^2$$

thus to calculate this we will need the quantity

\begin{eqnarray}
\frac{1}{2m}\frac{\partial^2F}{\partial k^2} + \frac{Dq}{2}\frac{\partial^2F}{\partial l^2} = \frac{1}{2m}\left(\frac{\partial^2F}{\partial k^2} + mDq\frac{\partial^2F}{\partial l^2}\right)
\end{eqnarray}

which - using the formulas (\ref{F_kk}) and (\ref{F_ll}) - will be

\begin{align}
\frac{1}{2m}\left(\frac{\partial^2F}{\partial k^2} + mDq\frac{\partial^2F}{\partial l^2}\right) = \frac{1}{2m}\left[\frac{A_r}{r} + A_{rr} + \frac{A_{\phi\phi}}{r} + A_r^2 + A_\phi^2\right]F.\label{harm_osc_en}
\end{align}

First let us calculate the expectation value, which means evaluating this expression at $r = 0$ and for the sake of simplicity let us write only the terms which do not contain 
$B(r).$ Thus 

\begin{align}
\frac{1}{2m}\left(\frac{\partial^2F}{\partial k^2} + mDq\frac{\partial^2F}{\partial l^2}\right)\bigg|_{r = 0} = \frac{1}{2m}\left[\alpha\left(4\phi(\beta + 1) + \sin(2\phi)(\beta - 1)\right) + \tilde{F}\right]F
\end{align}

where $\tilde{F}$ contains terms which contain $B(r)$ and its derivatives.\\
The interesting results are obtained when one calculates the conditional probability of the energy. Indeed let us look at the cases where we evaluate (\ref{harm_osc_en}) at 
$l = 0$ (which means $\phi = (n + 1/2)\pi$) or at $k = 0$ (equivalent to $\phi = n\pi$):

\begin{align}
\frac{1}{2m}\left(\frac{\partial^2F}{\partial k^2} + mDq\frac{\partial^2F}{\partial l^2}\right)\bigg|_{\phi = n\pi} &= \frac{1}{2m}\left[\alpha\left(4n\pi(\beta + 1)\right) + \tilde{F}\right]F\label{Energy_HO_k}\\
\frac{1}{2m}\left(\frac{\partial^2F}{\partial k^2} + mDq\frac{\partial^2F}{\partial l^2}\right)\bigg|_{\phi = \left(n + \frac{1}{2}\right)\pi} &= \frac{1}{2m}\left[\alpha\left(4\left(n + \frac{1}{2}\right)\pi(\beta + 1)\right) + \tilde{F}\right]F\label{Energy_HO_l}
\end{align}

\item Let us substitute the values of $\beta$ and $\alpha$ into (\ref{Energy_HO_l}) and leave only the first term:

\begin{align}
\frac{1}{2m}\left(\frac{\partial^2F}{\partial k^2} \right. &+ \left. mDq\frac{\partial^2F}{\partial l^2}\right)\bigg|_{\phi = \left(n + \frac{1}{2}\right)\pi} = \frac{1}{2m}\left[\frac{1}{4K}\sqrt{\frac{m}{Dq}}\left(4\left(n + \frac{1}{2}\right)\pi\left(\frac{Dq}{m\lambda} + 1\right)\right)\right]F = \nonumber\\
&= \frac{\sqrt{m}(Dq + m\lambda)}{2Km^2\lambda\sqrt{Dq}}\left(n + \frac{1}{2}\right)\pi F = \frac{\sqrt{m}\left(\sqrt{Dq} + \frac{m\lambda}{\sqrt{Dq}}\right)}{2Km^2\lambda}\left(n + \frac{1}{2}\right)\pi F = \nonumber\\
&= \frac{\sqrt{\frac{Dq}{m}} + \sqrt{\frac{m}{Dq}}\lambda}{2Km\lambda}\left(n + \frac{1}{2}\right)\pi F\label{Energy_HO_substitute}
\end{align}
 
This expression is very similar to the energy eigenvalues for the harmonic oscillator in quantum mechanics. The similarities will be more transparent if we express the $\omega$ 
angular frequency with $D$. Since $E = -Dx$ the potential energy will be (note that this term is multiplied with $q$) $\frac{1}{2}Dqx^2.$ This should be equal to 
$\frac{1}{2}m\omega^2x^2$ thus 

$$\omega = \sqrt{\frac{Dq}{m}}.$$

Also we may notice the $1/(2Km\lambda)$ term which was identified at the infinite potential well problem as $\bar{h}.$ Substituting these into (\ref{Energy_HO_substitute}) we obtain

\begin{align}
\frac{1}{2m}\left(\frac{\partial^2F}{\partial k^2} + mDq\frac{\partial^2F}{\partial l^2}\right)\bigg|_{\phi = \left(n + \frac{1}{2}\right)\pi} = \bar{h}\left(\omega + \frac{\lambda}{\omega}\right)\left(n + \frac{1}{2}\right)\pi F\label{Energy_HO_final}
\end{align}

This is almost the same as the result obtained in quantum mechanics, the only difference is the $\lambda/\omega$ term. Thus the results of the harmonic oscillator are obtained 
if we take the limit $\lambda\to 0$ while $2Km\lambda$ remains constant.

\item The previous calculation is valid if $Dq > 0$ thus the electric field is an attracting field. If one substitutes $D = 0$ into (\ref{harmonic_char}) we do not obtain the 
result of the free particle. This is not surprising since the coordinate transformations from $(k,l)$ to $(r,\phi)$ are only valid if $D\neq 0.$ Also if $Dq < 0$ then $\phi$ 
becomes complex which means that instead of the $\sin$ and $\cos$ functions the $\sinh$ and $\cosh$ functions appear. This will change the behavior of the expectation values 
since in this case there will not be multiple points when either $l$ or $k$ is zero thus the discrete structure disappears.

\end{itemize}

\subsection{The double-slit experiment}

This model provides a very clear explanation of the results of the double-slit experiment. If we look at the arrangement of the experiment we may model the "box" with the 
two slits as a  electric field which is zero inside the box and at the slits and infinity at the walls. In this case the time independent equation takes the form

\begin{align}
0 &= \frac{1}{2K}\frac{\partial^2 f}{\partial p_1^2} + \frac{1}{2K}\frac{\partial^2 f}{\partial p_2^2} + \frac{1}{2Km^2\lambda}\frac{\partial^2 f}{\partial x_1^2} + \frac{1}{2Km^2\lambda}\frac{\partial^2 f}{\partial x_2^2} - \frac{p_1}{m}\frac{\partial f}{\partial x_1} - \frac{p_2}{m}\frac{\partial f}{\partial x_2} - \nonumber\\
&- qE_0(x_1)\frac{\partial f}{\partial p_1} - qE_0(x_2)\frac{\partial f}{\partial p_2}
\end{align}

where $f = f(x_1, p_1, x_2, p_2).$ Because of the symmetry of the potential changing the variables with index 1 and 2 will not change the differential equation which means 
$f(x_1, p_1, x_2, p_2) = f(x_2, p_2, x_1, p_1).$ Now if look at the solution of the infinite potential well problem the wave behavior is obvious. Now what happens if measure 
one particle? This will mean that e.g. the particle labeled with 1 will be in an electric field $E_0 + E_1$ but the other one will not. This breaks the symmetry and with the 
new electric field present $f(x_1, p_1, x_2, p_2) \neq f(x_2, p_2, x_1, p_1).$

\subsection{Measurement theory}

Since the first successes of quantum mechanics there has been a debate about a specific aspect of the interpretation and that is the role of measurements. The challenge is that 
in case of a measurement the wave function collapses to a specific state though it evolves deterministically through the Schrödinger-equation, containing all states.\\
In the model presented here this problem can easily be solved. If we think about it a measurement of a quantity means that the value of that quantity is fixed, which means that 
during a measurement it is not the total probability density function which is measured but only a conditional probability. For example if we measure the position $x$ of a system 
then it is not $f(x,p)$ but $f(x,p|x = X)$ is measured where $X$ is the obtained position of the system. If we look at the expression of the conditional probabilities and compare 
them to $f(x,p)$ we see that these describe different behaviors depending on what is the condition. This also explains why different types of measurements may lead to different 
types of results or interpretations.

\section{Summary and outlook}

In this paper the random field quantization method method is introduced which was motivated by that this maybe a good candidate for a model that explains quantum phenomena 
with classical statistical methods. It was shown that putting charged particles into a random electric field explains the processes of the double-slit experiment and the solve 
the problem of measurements but also gives back the discrete energy levels of the harmonic oscillator and the infinite potential well. Also the $n$-dependence ($n$ being the level) 
of these values also matches the one obtained in quantum mechanics ($n+\frac{1}{2}$ and $n^2$ respectively). During the construction of the model a constant with action dimension 
naturally emerged from the theory.\\
Of course there are a lot questions regarding this model which have to be answered in a satisfactory fashion.

\begin{itemize}
\item \underline{Positivity}: In all of the solutions there appear various constants or functions that are the coefficients of some basic solutions. Unfortunately the results 
for the probability density functions are too complex in either methods to check whether there exists coefficients for which the density function is indeed a probability density. 
\item \underline{The hydrogen atom}: The hydrogen atom is the simplest example in quantum mechanics that both can be solved exactly and the results are well measured through 
various experiments. It would be a good way to test the viability of this method to calculate the probability density function and the expectation value of the energy in this 
case as well.
\item \underline{The role of $\lambda$}: The $\lambda$ parameter was introduced to simplify the calculation, but in the end it not only ended up in the results of the calculations 
but it is an essential parameter when one wants to compare the results obtained via this method to the ones obtained in quantum mechanics. It would be desirable to investigate the 
physical meaning - if there is any - of this parameter. Also in the calculation it was put in as a constant but if look at the expression (\ref{path_density_mod}) carefully one 
can also replace the $\lambda$ constant with a $\lambda(t,\underline{x}, \dot{\underline{x}})$ function provided that this function is always bounded and positive.
\item \underline{The connection to quantum mechanics}: Though the discrete structure of the energy of the harmonic oscillator and the infinite potential well are similar to the 
results obtained in quantum mechanics, the legitimacy of the method would highly increase if one would provide a direct relationship between the wave function in quantum mechanics 
and the probability density function in \ref{diff_final}. One of the possible candidates could be the Wigner-function. In \cite{Wigner_function} the authors provide a differential 
equation for the Wigner-function of the damped Harmonic oscillator (equation 3.63.):

\begin{align}
\frac{\partial f(x,y,t)}{\partial t} &= -\frac{y}{m}\frac{\partial f(x,y,t)}{\partial x} + m\omega^2x\frac{\partial f(x,y,t)}{\partial y} + (\lambda - \mu)\frac{\partial [xf(x,y,t)]}{\partial x} + \nonumber\\
&+ (\lambda + \mu)\frac{\partial [yf(x,y,t)]}{\partial y} + D_{qq}\frac{\partial^2 f(x,y,t)}{\partial x^2} + D_{pp}\frac{\partial^2 f(x,y,t)}{\partial y^2} +\nonumber\\
&+ 2D_{pq}\frac{\partial^2 f(x,y,t)}{\partial x\partial y}\label{harm_osc_Wigner}
\end{align}

where $y$ is the momentum, $\lambda$ is the friction coefficient, $\mu$ is the coupling coefficient and the matrix elements $D_{pp},\ D_{pq},\ D_{qq}$ are the diffusion 
coefficients. If we compare this with the differential equation for the harmonic oscillator obtained in this model, that is

\begin{eqnarray}
0 = \frac{1}{2K}\frac{\partial^2 f}{\partial p^2} + \frac{1}{2Km^2\lambda}\frac{\partial^2 f}{\partial x^2} - \frac{p}{m}\frac{\partial f}{\partial x} + qDx\frac{\partial f}{\partial p},\label{basiceq2d_harmo_osc}
\end{eqnarray}

we find that the two differential equations are identical if in (\ref{harm_osc_Wigner}) we substitute 
$\lambda = 0,\ \mu = 0,\ \partial f/\partial t = 0,\ D_{pq}=0,\ D_{qq} = \frac{1}{2Km^2\lambda},\ D_{pp} = \frac{1}{2K}.$ Of course this does not mean that the distribution density 
is necessarily the Wigner-function but further investigations in this direction might be promising.
\end{itemize}

Even if these questions are answered there are interesting questions that can be asked regarding this approach.

\begin{itemize}
\item \underline{Field theory}: To simplify the problem we used the Lorentz-force for a single particle. However it can also be written for a continuous $\rho$ electric charge 
density and an $\underline{j}$ electric current density. In this case if one calculates the probability functional $f$ it will depend on the functions $\rho(\underline{x}, t)$ 
and $\underline{j}(\underline{x}, t).$ It would be interesting to derive how $f$ is actually depend on $\rho$ and $\underline{j}$ and what kind of (functional) differential 
equation can be set up for it.

\item \underline{Including the magnetic field}: For simplicity we have only looked at cases where only a random electric field is present. It would be interesting to see how the 
inclusion of the random magnetic field would change the formulas for the probability density function (the difficulty is that if both the random electric and magnetic fields 
are present then expressing these fields in terms of $\underline{x},\  \dot{\underline{x}},\ \ddot{\underline{x}}$ from the Lorentz-force is ambiguous).

\item \underline{Relativistic theory}: We have shown that if we consider an $M$ four dimensional region of the space-time the probability density functional including the 
electric and magnetic random fields will be 

$$f(\underline{E},\underline{B})=\exp\left(-\frac{K}{2}\int_M d^3xdt\left(\frac{\epsilon_0}{2}\underline{E}^2 + \frac{1}{2\mu_0}\underline{B}^2\right)\right).$$

We may notice that the integrand is $T_{00},$ the zero-zero component of the stress energy tensor. Since in relativistic notation $d^3xdt = d^4x$ the above integral may be 
rewritten as 

$$f(F_{\lambda\delta})=\exp\left(-\frac{K}{2}\int_M d^4 T_{\mu\nu}h_{\mu\nu}\right),$$

where $h_{\mu\nu}$ is a symmetric tensor such that the integrand is positive definite. The Lorentz-force in this case will be of the form 

$$\frac{dp^a}{d\tau} = q(F^{ab} + \tilde{F}^{ab})u_b$$

where $p^a$ is the four momentum, $\tau$ is the proper time, $u_b$ is the four velocity, $F^{ab}$ is the electromagnetic tensor of the given electromagnetic field and 
$\tilde{F}^{ab}$ is the electromagnetic tensor of the random field. The difficulty in this case is to find a probability measure which is covariant and is constructed from the 
fields $A_\mu.$

\item \underline{Constraints on the random field}: The distribution functions calculated are obtained with the presumption that the fields of the surrounding particles can 
be arbitrary. This is a simplification since all of these fields have to obey Maxwell's equations. What is neglected here are the equations that are independent of the charge 
and current distributions, specifically

\begin{align}
div\underline{B} &= 0\nonumber\\
rot\underline{E} &= -\frac{\partial\underline{B}}{\partial t}.\nonumber
\end{align}

It would be interesting to examine how the distribution function of these fields change if we incorporate these constraints as well.

\item \underline{Other interactions} In this paper only the electromagnetic field was considered but it would be interesting to see whether a similar approach can be done in 
the case of the strong and the weak interactions. The difficulty is that neither of these interactions have a classical version.

\end{itemize}

\section{Acknowledgments}

I would like to thank Mih\'aly Csirik and G\'abor Homa for their valuable help and insights.

\end{document}